\documentclass[epj]{webofc}
\usepackage[varg]{txfonts}   
\usepackage{graphicx}
\usepackage{amsmath}
\usepackage{mathabx} 
\woctitle{Hot Planets and Cool Stars }
\begin{document}
\title{Small flow rate can supply inwardly migrating shortest-period planets
}
%
%

\author{Stuart F. Taylor\inst{1,2}\fnsep\thanks{\email{astrostuart@gmail.com
    }} 
}

\institute{
Participation Worldscope/Global Telescope Science\\
Hong Kong, SAR China
and Sedona, Arizona, U.S.A. 
\and Job Seeking
          }

\abstract{%
The number of exoplanets found with periods as short as one day and less
was surprising given how fast these planets had been expected to 
migrate into the star due to the tides raised on the star 
by planets at such close distances. It has been seen as improbable that
we would find planets in such a small final fraction of their 
lives \cite{ham09}.
The favored solution has been that the tidal dissipation is much
weaker than expected, which would mean that the final infall would
be a larger fraction of the planets' life. 
We find no reason, however, to exclude the explanation that a small
number of planets are continuously sent migrating inwards such
that these planets indeed are in the last fraction of their lives.
Following the observation that the distribution of medium planets 
disfavors tidal dissipation being significantly weaker than has been
found from observations of binary stars~\cite{tay12a}, 
we now show that the numbers of planets in such a ``flow''
of excess planets migrating inwards is low enough that even
depletion of the three-day pileup is a plausible source.
Then the shortest period occurrence distribution would be shaped by 
planets continuously being sent into the star, which may explain
the depletion of the pileup in the Kepler field relative to the
solar neighborhood \cite{how12}. 
Because Kepler observes above the galactic plane, 
\cite{how12} suggested the Kepler field 
may include an older population of stars. 
The tidal dissipation strength in stars due to giant planets
may be not greatly weaker than it is in binary stars.

}
\maketitle
\section{Introduction}
The discovery of too many planets found in the last fraction of their 
lifetimes is among the most unexplained exoplanet discoveries \cite{ham09}.
Either tidal migration is slower than expected,
or these planets have recently been sent inward.
The favored explanation has been that tidal dissipation
within the star from tides on the star raised by these planets
is much weaker than the tidal dissipation 
as measured from binary star statistics \cite{ham09}, 
but we show the more plausible explanation is 
that more planets migrate inwards than expected.
It is thought that inwardly migrating
planets circularize and nearly cease inward migration,
avoiding infall into the star.
We propose that planets in the pileup may be caused to migrate more
rapidly than expected 
than if most orbits are circular.

We show that the discrepancy between the inferred tidal strength found
by \cite{tay12a,tay12b} is also supported by directly calculating
the infall times for the Kepler planet candidates (hereafter ``planets'').

\label{intro}

\begin{figure}
\centering
\includegraphics[width=8cm,clip]{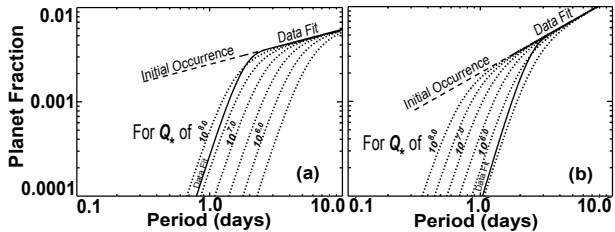}
\caption{Migrated evolution compared to data for masses and radii for
    an example age of 4.5 Gyr for planets with masses
    summed over the ranges given in the text, 
    (a) for ``large'' planets, and  
    (b) for ``medium'' planets.
}
\label{fig-distLMo-c}       
\end{figure}

\section{Inward migration as source of shortest period planets}
\label{sec-inward}                 
The strength of dissipation had been expected to be stronger than
a tidal dissipation strength ``$Q^{\prime}_{\ast}$'' of $10^7$, 
which is expressed as $Q^{\prime}_{\ast}$ being ``less than'' $10^7$, 
where the dissipation strength is proportional to $1/{Q^{\prime}_{\ast}}$.
However, careful studies based on the evolution of the 
occurrence distribution have presented planet
distribution to be 
consistent with ${Q^{\prime}_{\ast}}$ weaker than $10^7$ 
(e.g., \cite{pen12}), when assuming no input of new planets
migrating inwards.
We find that the rate of planet migrating inwards
need only be on the order of $10^{-12}$ or less planets per year in order
for the presence of the planets with these shortest periods
to be consistent with a tidal dissipation $Q^{\prime}_{\ast}$ strenght of 
$10^{7}$ or stronger. 
This number is  low enough to  be supplied
by a decay in the long observed three day pileup of giant planets,
or it could be due to a hypothesized flow of planets from further out.
Both explanations have  support:
The number of planets in the pileup of planet candidates 
(hereafter ``planets'') found by Kepler has been reported
to be 40\% lower than for the planets found in the solar neighborhood
by \cite{how12},
who suggest that perhaps the Kepler field, being above the
galactic plane, has an older population of stars.
The pileup may have decayed.
The hypothesis of a flow of ``high eccentricity migration'' 
(HEM) of giant planets 
supplying the pileup has been suggested by
\cite{soc12}, though such eccentric planets 
have not yet been found  \cite{daw12}. 

The shortest period distribution of planets has a fall off
that within large uncertainty 
is consistent with migration due to tides in the star 
caused by the planet \cite{tay12a}.
Tidal migration due to these ``stellar'' tides  produce an occurrence
distribution with a fall off that has a power index of 13/3 
based on  equations of \cite{jac09}.
The two-power law fit to the Kepler planet occurrence
distribution by \cite{how12} gives a fall off in the
shortest period region                 
that for giant and medium radii planets is reasonably close to 13/3.
We obtain the power index of the fall off
from the fit by \cite{how12}, 
by suming their two indices $\beta + \gamma$,
which is the power index of their 
fit function in the limit of the period going to zero.
We call the three ranges in radii used by \cite{how12} 
large, medium and (relatively) small.
We assign the following radii and mass ranges to use in 
calculations of tidal migration,
in earth radii $R_\Earth$, and earth masses, $M_\Earth$:
``Large'' planets of radii from 8 to 16 $R_\Earth$ 
  and of mass from $100$ to $2000 M_\Earth$,
``medium'' planets of radii 4 to 8 $R_\Earth$ 
    and of mass from $100$ to $2000 M_\Earth$,
and ``small'' planets of radii 2 to 4 $R_\Earth$ 
    and of mass from $10$ to $100 M_\Earth$.
The power indices found by \cite{how12}, for giant planets, $4.5 \pm 2.5$,
and for medium planets, $4.8 \pm 1.3$, 
are both consistent with the value of 13/3 
indicating currently inwardly tidally migrating planets,
but the power index for superearth planets, $2.9 \pm 0.4$, is too low
to be from ongoing tidal migration. The slope 
of the superearth planets likely originated with their formation.

We present the evolution of the fall off as a function of $Q^{\prime}_{\ast}$ 
for a summed range of stellar ages in figure~\ref{fig-distLMo-c}.
We calculate tidal migration operating on an initial occurrence distribution 
of a single power law to produce a distribution with a fall off, giving 
the two-power law distribution that resembles the actual distribution. 
The result, as shown in figure~\ref{fig-distLMo-c}
is similar to the two-power law distribution of \cite{how12}. 


In figure~\ref{fig-distLMo-c}, 
we compare fits by \cite{how12} 
with fall offs calculated for a range of tidal distributions.  
We show the occurrence distributions of \cite{how12} for  
giant and medium radii planets
in figure~\ref{fig-distLMo-c}, 
plotted against our calculations of 
occurrence distributions summed for mass
after a representative migration time of 4.5 Gyr,
shown for several values of tidal dissipation
strengths $Q^{\prime}_{\ast}$.
The fall off for giant and medium planets could 
be interpreted as giving different values of
$Q^{\prime}_{\ast}$ for differently sized planets,
but it could also be explained by more giant planets migrating inwards. 
The difference appears to correspond with the radii
where the pile up of
giant planets occurs, which suggests that the pile up
is related to the higher 
occurrence of the shortest period giant planets.
This difference appears to be best reconciled
by a larger rate (flow) of planets migrating from further out.
This may or may not be related to a more distant
HEM flow of giant planets that 
\cite{soc12} 
propose may have supplied at least a part of the pileup,
because depletion of the pileup may be sufficient. 

\begin{figure}
\centering
\includegraphics[width=9cm,clip]{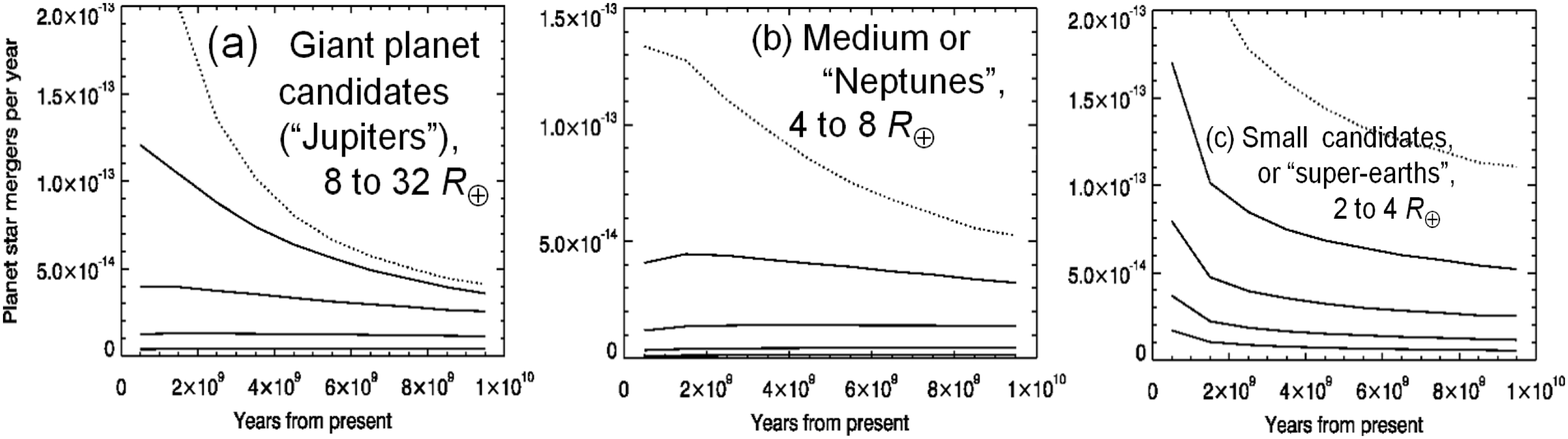}
\caption{Future fall in rates for the three ranges of planet radii 
  (panels for large, medium, and small radii) each with plots for tidal
  dissipation values of $\log(Q^{\prime}_{\ast})$ 
  values of $10^{6.5}$ (dotted, top), 
  $10^{7.0}$, $10^{7.5}$, $10^{8.0}$, and $10^{8.5}$.
  Rate of fall in calculated based on fit.}
\label{fig-rateforfit}
\end{figure}

\begin{figure}
\centering
\includegraphics[width=9cm,trim=0cm 0cm 0cm 0.15cm,clip=true]{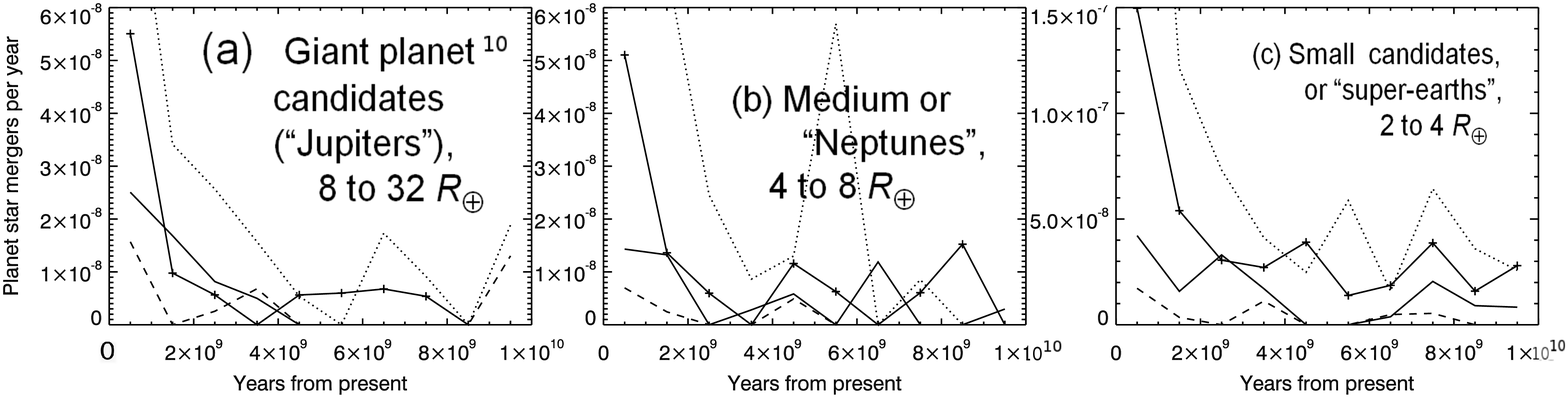}
\caption{Rate of fall in calculated by modeling when Kepler candidates
  would fall in, with the same range of values as in 
  figure~\ref{fig-rateforfit}.
  }
\label{fig-ratefordata} 
\end{figure}

In the three panels of figure~\ref{fig-rateforfit} 
we show for the three size ranges of planets
the calculated future infall rates plotted for 
stellar tidal dissipation values of ${Q^{\prime}_{\ast}}$
from $10^{6.5}$ to $10^{8.5}$, using the fits of
\cite{how12} as the initial occurrence distributions. 
We also show infall calculated directly using data directly in 
figure~\ref{fig-ratefordata}, where we see the same
discrepancy, though the result is noisy.
The correct value of $Q^{\prime}_{\ast}$ is the one
that gives infall rates
that decrease at a rate no faster than 
the supply of planets  decreases 
as stars age. 
The only way to maintain the presence of
more giant than medium planets in the shortest period range is 
if there is a larger inward flow of giant planets than for medium planets.
(This would only start adding to the infall after the new planets migrate in).
Arriving planets would produce a flatter curve for $Q^{\prime}_{\ast}$ 
closer to $10^7$. 

Inward planet migration might cause pollution of the star
if other planets orbits are disrupted.
The correlation between stellar Fe/H and eccentricity presented in
figure 3 of 
\cite{tay12b} suggests that high eccentricity planets
could be associated with recent pollution.
For systems found by radial velocity with periods less than 200 days
(exoplanets.org),
we find a less than 3\% chance that Fe/H values for 
stars hosting planets with orbits of eccentricity
above and below 0.35 represent the same population.

\begin{figure*}
\centering
\includegraphics[trim=0cm 0.2cm 0cm 10.6cm,width=5cm,clip=true]{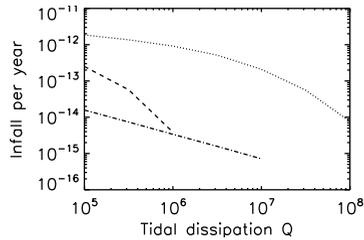} 
\caption{The difference in planet infall
rate per year as a function of $Q^{\prime}_{\ast}$ 
between the flow at 1 and 10 Gyr, shown for three radii
ranges. This shows that the likely rate of inward flow of giant planets (dotted)
is higher than that for medium (dashed) and smaller (dot-dashed) planets.
No line is shown for $Q^{\prime}_{\ast}$ values not needing a flow.
It takes only this rate of inward planet flow to maintain
a constant future infall rate.
}
\label{fig-rateVsQ}       
\end{figure*}

\section{Small supply required  -- pileup depletion probably sufficient}
\label{sec-pileup}


We find that the rate of planet infall required to provide the
shortest period planet population is small enough to reasonably
expect that  depletion of the pileup or a migration of planets
from further out could supply enough planets to explain
short period planets continually infalling.
We show the rates required in  figure~\ref{fig-rateVsQ}, 
where we have taken the difference in the rates of infall
(as in figure~\ref{fig-rateforfit})
at 1 and 10 Gyr.
The rate of this flow only needs to be less than $10^{-12}$ 
giant planets per year 
per star for the fall in rates to be consistent. 



The 40\% smaller size in the pileup in the Kepler field, \cite{how12}, 
compared to ground surveys that predominantly sample the solar neighborhood
 \cite{mar05}
could indicate that the pileup decreases with age.
This decrease is likely large enough to supply the small ongoing infall
of planets migrating into the star.
It is important to study whether planets in the pileup region might
be having their eccentricity pumped up, or might be disrupted from
planets migrating from further out, decreasing the numbers of 
planets in the pileup.

The first sign of stronger tidal migration could come in only 
a few years, when enough time has
elapsed to measure decreases in the shortest period planets \cite{ham09}. 
An important measure of the rate of 
planet infall will be observing transients from planet/star mergers,
which will likely have sufficient luminosity to be observable from nearby
galaxies by upcoming or current transient surveys
\cite{tay10,met12}.





\begin{thebibliography}{}
%
%

\bibitem[Hamilton(2009)]{ham09}
Hamilton, D.P., Nature \textbf{460}, 1086, (2009) 

\bibitem[Taylor(2012)]{tay12a} 
{Taylor, S.F.} arXiv:astro-ph/1206.1343 (2012a)

\bibitem[Howard et al.(2012)]{how12}
Howard, A.W., Marcy, G.W., \& Bryson, S.T. et al., ApJS \textbf{201}, 15 (2012)

\bibitem[Taylor(2012b)]{tay12b}
{Taylor, S.F.} arXiv:astro-ph/1211.1984 (2012b)

\bibitem[Penev et al.(2012)]{pen12}
Penev, K., Jackson, B., Spada, F. \& Thom, N.,    
Ap.J. \textbf{751}, 96 (2012) 

\bibitem[Socrates et al.(2012)]{soc12}
Socrates, A., Katz, B., Dong, S. \& Tremaine, S., ApJ \textbf{750}, 106 (2012)

\bibitem[Dawson et al.(2012)]{daw12}
Dawson, R.I., Murray-Clay, R.A., \& Johnson, J.A., arXiv:astro-ph/1211.0554	
(2012) 

\bibitem[Jackson et al.(2009)]{jac09}
Jackson, B., Barnes, R. \& Greenberg, R., ApJ, \textbf{698}, 1357 (2009)

\bibitem[Marcy et al.(2005)]{mar05}
Marcy, G., Butler, R. P., Fischer, D., et al., PTPS \textbf{158}, 24 (2005)

\bibitem[Taylor(2010)]{tay10}
{Taylor, S.F.} arXiv:1009.4221, (2010)  

\bibitem[Metzger(2012)]{met12}
Metzger, B.D., Giannios, D. \& Spiegel, D.S., MNRAS, \textbf{425}, 27 (2012)

\end{thebibliography}
%
%

\end{document}